\documentclass[prd,twocolumn,superscriptaddress,amsmath,amssymb,nofootinbib]{revtex4-1}

\begin{document}

\title{Role of symmetries in the Kerr-Schild derivation of the Kerr black hole}

\author{Eloy Ay\'on-Beato}
\email{ayon-beato-at-fis.cinvestav.mx}
\affiliation{Departamento de F\'{\i}sica, CINVESTAV--IPN, Apdo.
Postal 14--740, 07000, M\'exico~D.F., M\'exico}
\affiliation{Instituto de Ciencias F\'isicas y Matem\'aticas,
Universidad Austral de Chile, Casilla 567, Valdivia, Chile}

\author{Mokhtar Hassa\"ine}
\email{hassaine-at-inst-mat.utalca.cl} \affiliation{Instituto
de Matem\'atica y F\'isica, Universidad de Talca, Casilla 747,
Talca, Chile}

\author{Daniel Higuita-Borja}
\email{dhiguita-at-fis.cinvestav.mx} \affiliation{Departamento de
F\'{\i}sica, CINVESTAV--IPN, Apdo. Postal 14--740, 07000,
M\'exico~D.F., M\'exico} \affiliation{Instituto de Ciencias
F\'isicas y Matem\'aticas, Universidad Austral de Chile, Casilla 567, 
Valdivia, Chile}

\begin{abstract}
In this work we explore the consequences of considering from the very
beginning the stationary and axisymmetric properties of the Kerr black hole
as one attempts to derive this solution through the Kerr-Schild ansatz. The
first consequence is kinematical and is based on a new stationary and
axisymmetric version of the Kerr theorem that yields to the precise
shear-free and geodesic null congruence of flat spacetime characterizing the
Kerr solution. A straightforward advantage of this strategy is that now the parameter $a$ appears naturally 
as associated to the
conserved angular momentum of the geodesics due to axisymmetry. The second
consequence is dynamical and takes into account the circularity theorem. In
fact, a stationary-axisymmetric Kerr-Schild ansatz is in general
incompatible with the circularity property warranted by vacuum Einstein
equations unless the remaining angular dependence in the Kerr-Schild
profile appears fixed in a precise way. Thanks to these two ingredients, the
integration of the Einstein equations reduces to a simple ordinary
differential equation on the radial dependence, whose integration constant
is precisely the mass $m$. This derivation of the Kerr solution is simple
but rigorous, and it may be suitable for any textbook.
\end{abstract}

\maketitle

\section{Introduction\label{Sec:Intro}}

One of the most relevant and well-known predictions of general relativity is
the existence of black holes. Nevertheless, an outstanding but less
recognized prediction claims that if the $10^{20}$ black hole candidates in
the observable Universe are indeed black holes then all of them,
independent of their formation process, are described by a single exact
solution of Einstein field equations found by Kerr a half-century ago
\cite{Kerr:1963ud}. The Kerr solution describes a stationary and axisymmetric
black hole only characterized  by its mass and angular momentum. The
uniqueness proof of this solution was a \emph{tour de force} of mathematical
physics that took over 30 years and was dubbed in its beginning as the
``no-hair'' conjecture for black holes \cite{Heusler:1996}.

The highly nonlinear nature of the Einstein field equations makes difficult
any attempt to provide a straightforward or elementary derivation of the Kerr
solution. In spite of that, it will be desirable for methodological and
pedagogical reasons to have in hand a derivation as close as possible to an
elementary one. This is the motivation behind pursuing different criteria to
understand from first principles this solution and its generalizations. One
of these criteria is to look for a metric describing an ``exact
perturbation'' of flat spacetime that propagates along a null direction. This
is commonly known as the Kerr-Schild ansatz \cite{KerrSchild:1965}, and takes
the following form
\begin{equation}\label{eq:KerrSchild}
g_{\mu\nu}=\eta_{\mu\nu}+2Sl_{\mu}l_{\nu},
\end{equation}
where $\eta_{\mu\nu}$ is the flat Minkowski metric, $S$ is a scalar profile
function and $l_\mu$ is the tangent vector to a shear-free and geodesic null
congruence. The main merit and advantage of the Kerr-Schild ansatz lie in the
fact that the Einstein equations reduce to a linear system of equations
without any approximation. For vacuum and some specific types of matter, the
shear-free and geodesic character of the null congruence has been widely
discussed and justified in the current literature
\cite{KerrSchild:1965,Debney:1969zz,Stephani2003,Bini:2014nga}. A relevant
criterion is encoded by the so-called Kerr theorem
\cite{Debney:1969zz,Cox:1976,Stephani2003} which provides the most general
class of shear-free and geodesic null congruences in flat spacetime.
Nevertheless, the resulting class being given implicitly as a root of an
algebraic equation defined by an arbitrary complex function is quite general.
In fact, only a specific election of such equation including information
about the angular momentum leads to the stationary and axisymmetric
configuration known as the Kerr solution \cite{KerrSchild:1965}. However,
this election and the specific way of introducing the angular momentum, apart
from yielding the desired result, do not have any physical justifications.

As shown in the present work, this lack of justification can be circumvented
by adopting a different perspective. This will be done assuming beforehand
the relevant symmetries that the final state of the gravitational collapse is
supposed to enjoy. As a consequence, the proof of the stationary and
axisymmetric version of the Kerr theorem will be more involved than the
classical one \cite{Cox:1976}, but the advantage of such approach lies in the
fact that the degeneracy produced by the classical derivation will be
completely fixed. Indeed, in this case, it is possible to justify the
uniqueness of the specific shear-free and geodesic null congruence of flat
spacetime that characterizes the Kerr solution. In addition, the significance
of the angular momentum as the conserved charge of the geodesic motion associated
to the axial symmetry becomes evident and no artifact is needed. This will be
shown explicitly in Sec.~\ref{sec:StaAxi}. The second important consequence
has to do with the incompatibility between the circularity theorem that must
satisfy any vacuum stationary and axisymmetric spacetime
\cite{Papapetrou:1966zz,Kundt:1966zz,Carter:1969zz} and the Kerr-Schild
ansatz, which is manifestly noncircular by construction. In
Sec.~\ref{sec:CircCond} we show that this incompatibility is cured only if
the scalar profile $S$ of ansatz (\ref{eq:KerrSchild}) has a precise
dependence on the remaining angular coordinate. All these ingredients put
together allow us to reduce the Einstein equations to a trivial first-order
linear ordinary differential equation having the mass as the unique
integration constant. These arguments are extended trivially in
Sec.~\ref{sec:KN} to the charged case, and provide as well an elementary
derivation of the Kerr-Newman black hole.

\section{\label{sec:StaAxi}Stationary Axisymmetric Kerr Theorem}

We start by revising the classical Kerr theorem
\cite{Debney:1969zz,Cox:1976,Stephani2003}. It states that any
shear-free and geodesic null congruence in Minkowski space
\begin{equation}\label{eq:FlatNull}
ds_0^2=\eta_{\mu\nu}dx^{\mu}dx^{\nu}=-2dudv+2d\zeta d\bar{\zeta},
\end{equation}
is given by $l=dv$ or by
\begin{equation}\label{eq:Null}
l=du+Y\bar{Y}dv+\bar{Y}d\zeta +Yd\bar{\zeta},
\end{equation}
where $Y=Y(u,v,\zeta,\bar{\zeta})$ is a complex function implicitly defined
by
\begin{equation}\label{eq:KerrTheo}
F(Y,\bar{\zeta}Y+u,vY+\zeta)=0,
\end{equation}
with $F$ an arbitrary function in its three complex dependences.

The proof can be performed in two steps. The most general \emph{null} vector
field in Minkowski space with $l_u\ne0$ is given by (\ref{eq:Null}). The
\emph{geodesic} and \emph{shear-free} conditions over $l$ are equivalent in
this context to the equations \cite{Stephani2003}
\begin{subequations}\label{eq:GeoShear}
\begin{align}
\left(\partial_{\bar{\zeta}}-Y\partial_u\right)Y &=0,\label{eq:GeoShear1}\\
\left(\partial_v-Y\partial_\zeta\right)Y&=0.\label{eq:GeoShear2}
\end{align}
\end{subequations}
Thus, $Y$ is at the same time an invariant of both operators in
(\ref{eq:GeoShear}). First, the characteristic system related to
Eq.~(\ref{eq:GeoShear1}) is
\begin{equation}\label{eq:CharSys1}
\frac{d\bar{\zeta}}{1}=\frac{du}{-Y}=\frac{dv}{0}=\frac{d\zeta}{0},
\end{equation}
which implies that the general solution to Eq.~(\ref{eq:GeoShear1}) must be
necessarily a function of the three integration constants of this ordinary
system, giving $Y=Y(\bar{\zeta}Y+u,v,\zeta)$. The second step is taking into
account the last functional dependence in the characteristic system of the
second equation (\ref{eq:GeoShear2})
\begin{equation}\label{eq:CharSys2}
\frac{dv}{1}=\frac{d\zeta}{-Y}=\frac{d\left(\bar{\zeta}Y+u\right)}{0},
\end{equation}
meaning that $Y=Y(\bar{\zeta}Y+u,vY+\zeta)$ or, equivalently, the implicit
representation (\ref{eq:KerrTheo}). For $l_u=0$, interchanging the roles of
$u$ and $v$ in the above arguments we conclude that $l=dv$ in this case.

As said in the Introduction, it is a well-educated guess for the concrete
form of the function $F$ in (\ref{eq:KerrTheo}) that yields, after a highly
nontrivial integration process of the vacuum Einstein equations, to the
stationary and axisymmetric Kerr black hole
\cite{KerrSchild:1965,Debney:1969zz,Stephani2003,Bini:2014nga}. Here, we are
interested in exploring the consequences of assuming from the very beginning
the symmetries of the involved physical system without appealing to any other
assumption. This can be done by assuming that both the profile $S$ and the
shear-free and geodesic null vector $l$ are stationary and axisymmetric. In
other words, we are going to provide a stationary-axisymmetric version of the
Kerr theorem. In order to achieve this task and facilitate as possible the
proof, we first write the Minkowski flat spacetime in coordinates where the
stationary-axisymmetric character is manifest, i.e.\ in cylindrical
coordinates
\begin{equation}\label{eq:FlatCyl}
ds_0^2=-dt^2+d\rho^2+\rho^2d\phi^2+dz^2,
\end{equation}
where the stationary and axisymmetric isometries are realized by the Killing
fields $k=\partial_t$ and $m=\partial_\phi$, respectively. The tangent vector
to any congruence on flat Minkowski spacetime compatible with such symmetries
is written in cylindrical coordinates as
\begin{equation}
l=l_t(\rho,z)dt+l_\rho(\rho,z)d\rho+l_\phi(\rho,z)d\phi+l_z(\rho,z)dz.
\label{eq:lCyl}
\end{equation}
Imposing the geodesic condition, $l^{\mu}\nabla_{\mu}l_{\nu}=0$, we get
\begin{equation}\label{eq:GeoCyl}
(l_\rho\partial_\rho+l_z\partial_z)l_{\mu}=\frac{l_\phi^2}{\rho^3}
\delta_{\mu}^\rho.
\end{equation}
We notice from the identity
$l^\mu\nabla_\mu(l_{\nu}l^{\nu})=2l^{\nu}l^\mu\nabla_{\mu}l_{\nu}$ that the
geodesic equation for the component $l_\rho$ is satisfied for a null geodesic
if the geodesic equations for the remaining components are satisfied. The
remaining equations are just given by $l(l_{t})=l(l_{\phi})=l(l_{z})=0$,
which means that these components are invariants over the integral curves of
the vector field $l$ due to the translational isometries which are still
present in cylindrical coordinates \eqref{eq:FlatCyl}. However, their
determining equations have the common characteristic system
\begin{equation}\label{eq:Affine}
\frac{d\rho}{l_\rho}=\frac{dz}{l_z},
\end{equation}
having a single independent invariant. Consequently, all the other invariants
must be functions of the independent one that we choose to be the component
$l_z$. Additionally, since for null geodesics $l_t\neq0$, we can use the
scale invariance of the affine parametrizations of the geodesics to fix this
invariant component as $l_t=1$. In summary, the tangent vector to the most
general \emph{stationary-axisymmetric null geodesic} congruence in flat space
can be written as
\begin{subequations}\label{eq:NullCyl}
\begin{equation}
l=dt+l_{\rho}d\rho+l_\phi(l_z)d\phi+l_zdz,
\end{equation}
where the component $l_\rho=l_\rho(\rho,z)$ is determined from the quadratic
null condition
\begin{equation}\label{eq:NullCond}
l_\rho^2=1-\frac{l_\phi(l_z)^2}{\rho^2}-l_z^2,
\end{equation}
\end{subequations}
and the component $l_z=l_z(\rho,z)$ is an invariant of the geodesic motion,
$l(l_{z})=0$.

We shall now impose the \emph{shear-free} condition which means that the shear
tensor of the congruence denoted by $\sigma_{\mu\nu}$ vanishes. This latter is
defined as the traceless contribution from the symmetric part of the
covariant derivative, $\nabla_{\mu}l_{\nu}$, previously projected into the
two-dimensional spacelike sector orthogonal to the null vector $l$
\cite{Stephani2003}. Hence, it is effectively a two-dimensional matrix having
only two independent components. These two components can be read off from
the positive definite quantity
\begin{equation}\label{eq:Sigma2}
2\sigma^2\equiv\sigma_{\mu\nu}\sigma^{\mu\nu}
=\nabla_{(\mu}l_{\nu)}\nabla^{\mu}l^{\nu}
-\frac{1}{2}\left(\nabla_{\mu}l^{\mu}\right)^2.
\end{equation}
A direct calculation for the flat spacetime congruence \eqref{eq:NullCyl}
gives the sum of squares
\begin{align}
2\sigma^2={}&\frac{\left(1-l_z^2\right)\left(\rho^2-l_\phi^2\right)}
{2\rho^2l_z^2} \left\{ \left( \frac{\partial_\rho l_z}{1-l_z^2} + \frac{\rho
l_z}{\rho^2-l_\phi^2} \right)^2 \right. \nonumber \\
&+ \left. \frac{1}{\rho^2l_\rho^2} \left[
\partial_\rho l_z\frac{dl_\phi}{dl_z}
+ l_zl_\phi\left(
\frac{\partial_\rho l_z}{1-l_z^2}-\frac{\rho l_z}{\rho^2-l_\phi^2}
\right)
\right]^2
\right\}, \label{eq:Shear}
\end{align}
where we have replaced $\partial_z l_z$ in terms of $\partial_\rho l_z$ using
the invariant condition $l(l_z)=0$. The \emph{shear-free} condition is
equivalent to impose $\sigma^2=0$, which implies the vanishing of the two
squared quantities. Using both conditions, we get the following equation
\begin{equation}\label{eq:ODE(l[phi])}
\frac{dl_\phi\left(l_z\right)}{dl_z}=-\frac{2l_zl_\phi\left(l_z\right)}{1-l_z^2},
\end{equation}
whose solution is given by
\begin{equation}
l_\phi\left(l_z\right)=-a\left(1-l_z^2\right),
\label{eq:l[phi]}
\end{equation}
where $a$ is an integration constant modulating the conservation of the
angular momentum of the geodesic motion due to the presence of the
axisymmetry. We shall see later that this constant is just the celebrated
parameter describing the \emph{angular momentum} of the Kerr black hole. We
would like to emphasize the natural emergence of this constant through this
approach in contrast with the standard derivation. Substituting the
expression of the angular component in the vanishing of the first squared
term in \eqref{eq:Shear} and using $l(l_z)=0$, one yields to a system for the
undetermined component $l_z$ given by
\begin{subequations}\label{eq:l[z],rho-z}
\begin{align}
\partial_\rho l_z&=-\frac{\rho l_z\left(1-l_z^2\right)}
{\rho^2-a^2\left(1-l_z^2\right)^2},\label{eq:l[z],rho2}\\
\partial_z l_z&=\frac{\rho l_\rho\left(1-l_z^2\right)}
{\rho^2-a^2\left(1-l_z^2\right)^2}.\label{eq:l[z],z}
\end{align}
\end{subequations}
It is easy to see that the integrability condition $\partial_z\partial_\rho
l_z=\partial_\rho\partial_z l_z$ is satisfied in this case. Consequently,
Eq.~(\ref{eq:l[z],rho2}) can be cast into the form
\begin{equation}\label{eq:PDE(l[z])}
\partial_\rho\left(\frac{l_z^2\left[\rho^2-a^2\left(1-l_z^2\right)\right]}
{1-l_z^2}\right)=0,
\end{equation}
which in turn implies that
\begin{equation}\label{eq:f(z)Def}
\frac{l_z^2\left[ \rho^2-a^2\left( 1-l_z^2 \right) \right]}{1-l_z^2}=f(z).
\end{equation}
Deriving this expression with respect to $z$ and using Eq.~(\ref{eq:l[z],z})
restricts the above function to satisfy
\begin{equation}\label{eq:f(z)}
f'(z)^2=4f(z), \quad \Longrightarrow \quad
f(z)=(z-z_0)^2.
\end{equation}
Additionally, using the translation invariance of Minkowski spacetime
\eqref{eq:FlatCyl} along the symmetry axis $z$ one can choose the integration
constant to be zero, $z_0=0$. Hence, the relation \eqref{eq:f(z)Def}
completely determines the implicit dependence of the component $l_z$ in terms
of the coordinates $\rho$ and $z$. Nevertheless, it is more convenient to
represent the component $l_z=dz/dr$ in terms of the affine parameter $r$.
From the invariant character of $l_z$, their definition is straightforwardly
integrated as
\begin{equation}\label{eq:z/l[z]}
\frac{z}{l_z}=r-r_0,
\end{equation}
and, using the shift invariance of the affine parameter one can set $r_0=0$.
Combining together Eqs. (\ref{eq:f(z)Def})-(\ref{eq:z/l[z]}), the affine parameter is implicitly given by
\begin{subequations}
\begin{equation}\label{eq:Constraint}
\frac{\rho^2}{r^2+a^2}+\frac{z^2}{r^2}=1,
\end{equation}
and allows us to express the tangent vector to the most general stationary-axisymmetric shear-free and 
geodesic null congruences in flat Minkowski spacetime as
\begin{equation}\label{eq:lCylFinal}
l=dt+\frac{r\rho}{r^2+a^2}d\rho-\frac{a\rho^2}{r^2+a^2}d\phi+\frac{z}{r}dz.
\end{equation}
\end{subequations}
This is precisely the expression for the familiar congruence characterizing
the Kerr solution
\cite{KerrSchild:1965,Debney:1969zz,Stephani2003,Bini:2014nga}. We emphasize
that requiring stationarity and axisymmetry has allowed us to remove the
infinite degeneracy involved in the election of the function $F$ determining
all the congruences allowed by the general Kerr theorem in
(\ref{eq:KerrTheo}).

A better representation for the congruence is achieved by noticing that Eq. \eqref{eq:Constraint} describes 
a family of ellipsoids of
revolution. They become spheres of radius $r$ for $a=0$; accordingly, the
constant $a$ denotes the departure from sphericity of the family and each
ellipsoid is labeled by the affine parameter $r$. It is natural to consider
coordinates adapted to the ellipsoids by using the label $r$ and a pair of
angles parametrizing each ellipsoid. Since the angle $\phi$ naturally
defines the revolution around the symmetry axis, the second angle can be
defined in order to satisfy the restriction \eqref{eq:Constraint} choosing
$\rho=\sqrt{r^2+a^2}\sin\theta$ and $z=r\cos\theta$. In these
\emph{ellipsoidal coordinates}, the tangent vector to the congruence is
specified without any extra condition by
\begin{equation}\label{eq:lEllip}
l=dt+\frac{\Sigma dr}{r^2+a^2}-a\sin^2\theta d\phi,
\end{equation}
and the flat Minkowski spacetime now reads
\begin{equation}\label{eq:FlatEllip}
ds_0^2=-dt^2+\left(r^2+a^2\right)\sin^2\theta d\phi^2
+\frac{\Sigma dr^2}{r^2+a^2}+\Sigma d\theta^2,
\end{equation}
where $\Sigma=r^2+a^2\cos^2\theta$. The deduction of these celebrated
ellipsoidal coordinates was the main motivation in looking for a stationary-axisymmetric version of the Kerr 
theorem.

The consequence for the Kerr-Schild ansatz (\ref{eq:KerrSchild}) is that
its general stationary and axisymmetric version is naturally written
using ellipsoidal coordinates as
\begin{equation}\label{eq:MetricEllip}
ds^2=ds_0^2+2S(r,\theta)\!\left(dt+\frac{\Sigma dr}{r^2+a^2}
-a\sin^2\theta d\phi\right)^2,
\end{equation}
where only the profile $S(r,\theta)$ is to be determined. This is the
starting point to explore the existence of stationary-axisymmetric solutions
of the Einstein equations by means of a Kerr-Schild transformation. We show
in the next section that it is even possible to restrict more the ansatz by
fixing the angular dependence of the profile using some circularity
arguments.

\section{\label{sec:CircCond}The circularity theorem: Kerr black hole}

We start this section by reminding that any stationary-axisymmetric spacetime
with commuting Killing fields $k=\partial_t$ and $m=\partial_\phi$ fulfills
the following geometrical identities \cite{Heusler:1996}
\begin{equation}\label{eq:Circularity}
\begin{aligned}
d*\left(k\wedge m\wedge dk\right) &=
2*\left(k\wedge m\wedge R(k)\right),\\
d*\left(k\wedge m\wedge dm\right) &=
2*\left(k\wedge m\wedge R(m)\right),
\end{aligned}
\end{equation}
where the Killing fields are expressed as one-forms, $k=g_{\mu\nu}k^\nu
dx^\mu$ and $m=g_{\mu\nu}m^\nu dx^\nu$, and the Ricci one-forms are specified
by $R(k)=R_{\mu\nu}k^\nu dx^\mu$ and $R(m)=R_{\mu\nu}m^\nu dx^\mu$. These
identities are the base of the so-called circularity theorem
\cite{Papapetrou:1966zz,Kundt:1966zz,Carter:1969zz}; for vacuum solutions,
the left-hand sides of \eqref{eq:Circularity} vanish which in turn implies
that the differentiated expressions are constants and must vanish at the
symmetry axis where $m=0$. For the stationary-axisymmetric Kerr-Schild ansatz
\eqref{eq:MetricEllip}, we get
\begin{equation}\label{eq:PDE(S)}
\begin{alignedat}{2}
0&=*\left(k\wedge m\wedge dk\right)&&=
-\frac{2\sin\theta}{\Sigma}\partial_\theta\left[\Sigma S(r,\theta)\right],
\\
0&=*\left(k\wedge m\wedge dm\right)&&=
\frac{2a\sin^3\theta}{\Sigma}\partial_\theta\left[\Sigma S(r,\theta)\right].
\end{alignedat}
\end{equation}
The first equalities are the Frobenius integrability conditions defining the
circularity property, which warrants that the planes orthogonal to the
Killing vectors at any point are integrable to surfaces tangent to those
planes in the whole spacetime. Consequently, by choosing coordinates along
the Killing fields and on their orthogonal surfaces, the circular metrics
become block diagonal. The second equalities in \eqref{eq:PDE(S)} imply that
the stationary-axisymmetric Kerr-Schild ansatz is not circular by
construction unless the profile is appropriately restricted as\footnote{We
thank A.~Anabalon for helping us to elucidate this interesting feature of the
Kerr-Schild ansatz.}
\begin{equation}\label{eq:S}
S(r,\theta)=\frac{rM(r)}{\Sigma},
\end{equation}
where the angular dependence is dictated by the function $\Sigma$. Only for
the above family of profiles is it possible to transform to Boyer-Lindquist-like coordinates 
\cite{Boyer:1966qh}
\begin{equation}\label{eq:BLCoord}
\tilde{t}=t-\int\frac{2rM(r)}{\Delta}dr, \qquad
\tilde{\phi}=\phi-\int\frac{2arM(r)}{\Delta}dr,
\end{equation}
with $\Delta=r^2+a^2-2rM(r)$, where now the circularity is manifest
\begin{align}\label{eq:MetricBL}
ds^2={}&-\frac{\Delta}{\Sigma}\!
\left(d\tilde{t}-a\sin^2\theta d\tilde{\phi}\right)^2\nonumber\\
&+\frac{\sin^2\theta}{\Sigma}\!
\left(ad\tilde{t}-(r^2+a^2)d\tilde{\phi}\right)^2
+\frac{\Sigma dr^2}{\Delta}+\Sigma d\theta^2.
\end{align}

It only remains to fix the radial dependence encoded in the function $M(r)$.
This is easily done by inspecting the radial component of the \emph{vacuum}
Einstein equations in the Boyer-Lindquist-like coordinates
\begin{equation}\label{eq:rrEinstein}
{G_r}^r=-\frac{2r^2}{\Sigma^2}\frac{dM(r)}{dr}=0\,;
\end{equation}
hence, $M(r)=m$ where the constant $m$ is identified with the mass. The
remaining Einstein equations are identically satisfied. In summary, the unique vacuum spacetime that can be 
written as a stationary-axisymmetric Kerr-Schild transformation from flat Minkowski spacetime is given by the
metric (\ref{eq:MetricBL}) with $\Delta=r^2+a^2-2mr$. This is precisely the
Kerr black hole \cite{Kerr:1963ud}.

\section{\label{sec:KN}Kerr-Newman black hole}

In this section, we extend the previous analysis to the electrovac case
following basically the same strategy. First, using Einstein equations with
the Maxwell energy-momentum tensor, the identities \eqref{eq:Circularity} are
now given by \cite{Heusler:1996}
\begin{equation}\label{eq:MaxwellCircular}
\begin{aligned}
-\frac14d*\left(k\!\wedge\!m\!\wedge\!dk\right) &=
*F(k,m)E_k+F(k,m)B_k,\\
-\frac14d*\left(k\!\wedge\!m\!\wedge\!dm\right) &=
*F(k,m)E_m+F(k,m)B_m,
\end{aligned}
\end{equation}
where the electric and magnetic one-forms, $E_X=-i_XF$ and $B_X=i_X\!*\!F$,
are defined in terms of the electromagnetic field strength
$F=dA=\frac12F_{\mu\nu}dx^{\mu}{\wedge}dx^{\nu}$ for any Killing field $X$.
Second, for a stationary-axisymmetric electromagnetic field,
$\pounds_kF=0=\pounds_mF$, it follows from the Maxwell equations that
$F(k,m)=0=*F(k,m)$ \cite{Heusler:1996}. Consequently, the right-hand sides of
\eqref{eq:MaxwellCircular} vanish leading to the electrovac circularity
\cite{Carter:1969zz}, and in the case of a stationary-axisymmetric
Kerr-Schild ansatz to the conditions \eqref{eq:PDE(S)} yielding as before to
the metric profile \eqref{eq:S}. The extension of the Kerr-Schild ansatz to
the electromagnetic field is done by considering the vector potential $A$
proportional to the shear-free and geodesic null vector $l$. As it happens
with the metric, a stationary-axisymmetric ansatz $A=\hat{S}(r,\theta)l$
determined by \eqref{eq:lEllip} is not circular by construction, since
$0=*F(k,m)\propto\partial_\theta[\Sigma\hat{S}(r,\theta)]$, unless the
electromagnetic profile has an angular dependence dictated by
$\hat{S}(r,\theta)=rQ(r)/\Sigma$. The remaining Maxwell equations impose that
the radial function is a constant $Q(r)=q$ identified with the electric
charge. The radial dependence of the metric is again determined from the
radial Einstein equation which is now given by ${dM(r)}/{dr}={q^2}/{2r^2}$,
and whose solution reads $M(r)=m-q^2/2r$ where the constant $m$ represents
the mass as before. The full Einstein-Maxwell system is fulfilled in this
way; consequently, the electrovac spacetime that can be written as a
stationary-axisymmetric Kerr-Schild transformation from the flat Minkowski
spacetime is given by the metric (\ref{eq:MetricBL}) with
$\Delta=r^2+a^2-2mr+q^2$ and the electromagnetic field $A=qrl/\Sigma$. This
is nothing but the so-called Kerr-Newman black hole \cite{Newman:1965my}.

\section{\label{sec:Sum}Conclusions}

The main objective of the present paper was to emphasize the crucial
importance that the symmetries can play in the Kerr-Schild derivation of the
Kerr solution in general relativity. Indeed, we have explicitly shown that
the assumption of stationarity and axisymmetry allows us to define in a unique
way the shear-free and geodesic null congruence of flat spacetime
characterizing the Kerr black hole. The advantage of this symmetric version
of the Kerr theorem lies in the fact that the infinite degeneracy present in
the original Kerr theorem is no longer present. Another advantage of the
process described here is that the meaning of the angular momentum as a
conserved quantity of the geodesic motion is revealed without any artifact.
The next step of the strategy consisted of using the circularity theorem
from which we have managed to integrate out the angular dependence of the
profile function. Finally, the Einstein equations reduce to an elementary
first-order differential equation in the Boyer-Linquidst-like coordinates
that is also trivially integrated. We believe that all these ingredients put
together constitute a straightforward, elementary, but rigorous derivation of
the Kerr black hole. A generalization in order to include the electric charge
can be carried out following the same lines yielding the Kerr-Newman
generalization.

As a natural extension of the present procedure, it will be interesting to
establish the equivalent of the stationary and axisymmetric Kerr theorem in
the (anti-)de Sitter spacetime, putting on a firmer ground the use of the Kerr-Schild
ansatz in this context. Another nontrivial task to explore is the extension
to higher dimensions where it is still not clear the role played by the
shear-free property.

\begin{acknowledgments}
We are thankful to A.~Anabalon for enlightening discussions. This work has
been funded by Grants No. 175993, No. 178346 and No. 243342 from CONACyT, together with
Grants No. 1121031, No. 1130423 and No. 1141073 from FONDECYT and Grant No. DPI20140053 from CONICYT - 
Research Council UK - RCUK. E.A.B. was partially supported by the ``Programa
Atracci\'{o}n de Capital Humano Avanzado del Extranjero, MEC'' from CONICYT.
D.H.B. was partially supported by the ``Programa de Becas Mixtas'' from CONACyT.
\end{acknowledgments}

\end{document}